\documentclass{elsart}
\usepackage{graphicx}

\newcommand{\kv}{\mbox{\boldmath$k$}}
\newcommand{\rv}{\mbox{\boldmath$r$}}

\begin{document}

\begin{frontmatter}

\title{Exact results on spin glass models}
\author{Hidetoshi Nishimori}
\address{Department of Physics, Tokyo Institute of Technology,
Oh-okayama, Meguro-ku, Tokyo 152-8551, Japan}

\begin{abstract}
Exact and/or rigorous results are reviewed for the Ising and $XY$/Villain
spin glasses in finite dimensions, such as the exact energy, correlation
identities and a functional relation between the distribution functions
of ferromagnetic and spin glass order parameters.
This last relation is useful to prove that
the phase space is not complex on a line in the phase diagram.
The spin wave theory neglecting periodicity is shown to give exact
results for the Villain model on a line in the phase diagram.
Implications of this fact
on the renormalization-group results for the random-phase
$XY$/Villain model in two dimensions are discussed.

\end{abstract}

\begin{keyword}
Spin glass \sep Ising model \sep random-phase $XY$ model
\sep Villain model \sep spin wave theory \sep gauge theory
\end{keyword}

\end{frontmatter}

\section{Introduction}
The mean-field theory of spin glasses is now well established.
The spin glass phase that exists at low temperatures is known to be
characterized by a complex phase space and slow dynamics.
It is a target of active current investigations to settle
whether or not the predictions of the mean-field theory apply
to realistic finite-dimensional systems.
It is generally difficult to clarify the properties
of finite-dimensional systems by analytical methods,
and numerical investigations are the main tools of
research.
The purpose of the present contribution is to overview the
available exact/rigorous results on spin glass models in
finite dimensions with several new additions to the list.
We also prove coincidence of our exact results with the
consequences of the spin wave theory for the random-phase
$XY$/Villain model.
This fact may shed a new light on the significance of existing
renormalization group arguments for the two-dimensional
random-phase $XY$/Villain model.

\section{Ising spin glass}

Let us first consider the Ising spin glass described by the Hamiltonian
  \begin{equation}
    H=-\sum_{\langle ij \rangle} J_{ij} S_i S_j,
   \label{Hamiltonian}
  \end{equation}
where $S_i=\pm 1$ and $J_{ij}=\pm J$, namely the $\pm J$ model.
The Gaussian model can be treated similarly.
This Hamiltonian is invariant under the gauge transformation
  \begin{equation}
     S_i\to S_i \sigma_i,~~~J_{ij}\to J_{ij}\sigma_i \sigma_j,
    \label{gauge-Ising}
   \end{equation}
where $\sigma_i $ is another Ising variable fixed arbitrarily
at each site.
Using this gauge invariance, we can derive many exact or rigorous
results \cite{Nishimori81,Nishimori_OUP},
some of which are listed below.
Unless stated otherwise, the only condition for the following
results to hold is that $K=K_p$, where $K=\beta J(=J/k_BT)$ and
$K_p$ is a function of $p$ (the probability that the quenched
random variable $J_{ij}$ is equal to $J$), $\exp (-2K_p)=(1-p)/p$.
No restrictions exist on the lattice structure or the spatial dimensionality.

The average energy is obtained exactly as
\begin{equation}
  [E]=-N_B J \tanh K
   \label{Ising-E}
\end{equation}
and the upper bound on the specific heat is
  \begin{equation}
    k_B T^2 [C] \le J^2 N_B {\rm sech}^2 K,
     \label{Ising-C}
  \end{equation}
where $N_B$ is the number of bonds in the system.
The square brackets denote the configurational average.
The gauge-invariant correlation function decays exponentially
  \begin{equation}
    [\langle S_0 \cdot\tau_{01}\tau_{12}\cdots \tau_{r-1, r}\cdot S_r \rangle ]
     = \tanh^r K,
     \label{Ising-gauge-correlation}
  \end{equation}
where $\tau_{j,j+1}(=\pm 1)$ is the sign of $J_{j, j+1}$,
and the pairs of neighbouring sites $(0,1), (1,2),\cdots ,(r-1,r)$
connect sites 0 and $r$.
The configurational average of the inverse correlation function is
identically unity
 \begin{equation}
   \left[ \frac{1}{\langle S_i S_j S_k \cdots \rangle} \right] =1,
    \label{Ising-inverse-correlation}
  \end{equation}
and the correlation function satisfies
 \begin{equation}
    [\langle S_i S_j \rangle ] =
      [\langle S_i S_j \rangle^2 ].
    \label{Ising-correlation}
  \end{equation}
By taking the thermodynamic limit in the above correlation identity
(\ref{Ising-correlation})
and then considering the limit of infinite separation of the two
sites $i$ and $j$, we conclude that the ferromagnetic order parameter
is equal to the spin glass order parameter, $m=q$.

\begin{figure}
  \begin{center}
   \includegraphics[width=.35\linewidth]{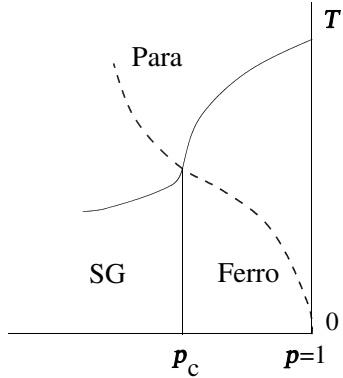}
  \end{center}
  \caption{Phase diagram of the $\pm J$ model and the
   Nishimori line (dashed).}
  \label{fig:phase_diagram1}
\end{figure}

All these results suggest a very special role played by the line
defined by $K=K_p$ on the phase diagram (Fig. \ref{fig:phase_diagram1}),
the Nishimori line.
In particular, the relation $m=q$ together with an inequality on the sign
of local magnetization \cite{Nishimori93,Nishimori_OUP}
 \begin{equation}
    M(K, K_p) =\left[ {\rm sgn}\langle S_i \rangle \right]
     \le M(K_p, K_p)
 \end{equation}
implies that the Nishimori line
marks a crossover between the purely ferromagnetic region
(where $m>q$ and $M$ is a decreasing function of $T$)
and a randomness-dominated region
(where $m<q$ and $M$ is an increasing function of $T$)
within the ferromagnetic phase.

An important generalization of the relation $m=q$ is the functional
identity $P_m(x)=P_q(x)$ under the same condition
$K=K_p$ \cite{Nishimori-Sherrington,Nishimori_OUP}.
Here, $P_m(x)$ is the distribution function of magnetization
\begin{equation}
 P_m(x)=\left[ \frac{\sum_{S}\delta \left(x-\frac{1}{N}\sum_i S_i \right)
  e^{-\beta H}}{\sum_{S}e^{-\beta H}}\right]
  \label{Ising-Pm}
\end{equation}
and $P_q(x)$ is the distribution of spin glass order parameter
defined in terms of two real replicas
\begin{equation}
 P_q(x)=\left[ \frac{\sum_{S}\sum_{\sigma}
    \delta \left(x-\frac{1}{N}\sum_i S_i \sigma_i \right)
  e^{-\beta H(S)-\beta H(\sigma )}}
    {\sum_{S}\sum_{\sigma}e^{-\beta H(S)-\beta H(\sigma )}}
   \right].
   \label{Ising-Pq}
\end{equation}
The proof of the identity $P_m(x)=P_q(x)$ is relatively straightforward if we
apply the gauge transformation to the expression (\ref{Ising-Pm}) and
use the condition.

Since the distribution function of magnetization $P_m(x)$
is always composed of at most two simple delta functions located at $\pm m$,
the identity $P_m(x)=P_q(x)$ implies that the distribution of the
spin glass order parameter $P_q(x)$ also has the same simple structure.
This excludes the possibility that a complex structure of the phase space,
which should be reflected to a non-trivial functional form
of $P_q(x)$, exists on the Nishimori line $K=K_p$.
Thus a spin glass (or mixed ferromagnetic) phase of the mean-field type,
if it exists in finite dimensions, should lie away from the line
in the phase diagram.

\section{Random-phase $XY$ and Villain models}
The same type of argument applies to the random-phase $XY$ model
(gauge glass)
 \begin{equation}
    H=-J \sum_{\langle ij\rangle } \cos
    (\theta_i-\theta_j -A_{ij} ) \label{XY-Hamiltonian}
 \end{equation}
with the phase distribution
\begin{equation}
   P(A_{ij})\propto e^{K_p \cos A_{ij}},
 \end{equation}
and the random-phase Villain (or periodic Gaussian)
model with nearest-neighbour Boltzmann factor
 \begin{equation}
  e^{V(\theta_i-\theta_j -A_{ij} )}=\sum_{m=-\infty}^{\infty}
    e^{-K(\theta_i-\theta_j -A_{ij} -2\pi m)^2/2},
          \label{Villain-BF}
  \end{equation}
where the phase variable obeys
 \begin{equation}
    P(A_{ij})\propto \sum_{m=-\infty}^{\infty}
    e^{-K_p(A_{ij} -2\pi m)^2/2}.
    \label{Aij-distribution-V}
 \end{equation}
We note here that the periodic distribution (\ref{Aij-distribution-V})
is equivalent to a non-periodic Gaussian distribution 
(often used in the literature) because
(\ref{Villain-BF}) is periodic in $A_{ij}$.
The Villain model (\ref{Villain-BF}) is often used as a
low-temperature effective model of the $XY$ model (\ref{XY-Hamiltonian}),
in particular in the analysis of the
Kosterlitz-Thouless transition in two dimensions.
We shall see that the results for the Villain model indeed
agree with those for the $XY$ model at sufficiently low temperatures.

These models are invariant under the gauge transformation
 \begin{equation}
     \theta_i \to \theta_i+\phi_i,~~~A_{ij}\to A_{ij}+\phi_i-\phi_j.
  \end{equation}
We simply list the results for the random-phase $XY$ model
that have been obtained using gauge invariance
\cite{Nishimori81,Ozeki-Nishimori,Nishimori_OUP},
corresponding to (\ref{Ising-E}), (\ref{Ising-C}),
(\ref{Ising-gauge-correlation}), (\ref{Ising-inverse-correlation})
and (\ref{Ising-correlation}) for the Ising case,
 \begin{eqnarray}
    &&[E]=-N_B J \, \frac{I_1(K)}{I_0(K)}
                 \label{XY-E} \\
   && k_BT^2 [C] \le J^2 N_B \left( \frac{1}{2}+\frac{I_2(K)}{2I_0(K)}
   -\left(\frac{I_1(K)}{I_0(K)}\right)^2 \right)
                 \label{XY-C} \\
   && \left[ \langle e^{i (\theta_0-A_{01}-A_{12}-\cdots
   -A_{r,r-1}-\theta_r)} \rangle \right]
   =\left( \frac{I_1(K)}{I_0(K)} \right)^r 
                   \label{XY-gauge-correlation} \\
   &&  \left[ \frac{1}{\langle e^{i \sum_j a_j \theta_j}\rangle }
       \right] =1
                    \label{XY-inverse-correlation} \\
  && \left[ \langle e^{i(\theta_0 -\theta_r)} \rangle \right]
   =\left[ \left|\langle e^{i(\theta_0 -\theta_r)} \rangle \right|^2\right],
                    \label{XY-correlation}
 \end{eqnarray}
where $I_j(K)$ is the modified Bessel function and
the $a_j$ are arbitrary numbers.
All these equations hold on any lattice in any dimension as long as
the condition $K=K_p$ is satisfied.

Correlation functions of bond variables can also be evaluated
under the condition $K=K_p$:
  \begin{eqnarray}
    &&\left[ \langle \cos (\theta_0-\theta_1-A_{01} )
    \cos (\theta_r-\theta_{r+1}-A_{r,r+1} ) \rangle \right]
   =\left( \frac{I_1(K)}{I_0(K)}\right)^2
        \label{XY-cos-correlation} \\
  && \left[ \langle \sin (\theta_0-\theta_1-A_{01} )
    \sin (\theta_r-\theta_{r+1}-A_{r,r+1} ) \rangle \right]
   =0,  \label{XY-sin-correlation}
 \end{eqnarray}
where $(01)$ and $(r, r+1)$ are pairs of neighbouring sites.
The second identity for the sine correlation is particularly interesting
since it means the absence of spatial correlation between chirality
degrees of freedom \cite{Nishimori_OUP}.

An identity between the distribution functions of ferromagnetic
and spin glass order parameters can also be proved, $P_m(z)=P_q(z)$
for $K=K_p$,
where the distribution function of the ferromagnetic order parameter is
defined as
 \begin{equation}
    P_m(z)= \left[ \left\langle \delta \left( z-\frac{1}{N}
   \sum_j e^{i\theta_j}\right) \right\rangle \right]
 \end{equation}
and that of the spin glass order parameter is given using two replicas
as in (\ref{Ising-Pq})
 \begin{equation}
    P_q(z)= \left[ \left\langle \delta \left( z-\frac{1}{N}
   \sum_j e^{i(\theta_j-\phi_j)}\right) \right\rangle \right],
 \end{equation}
where two replicas of the same system have dynamical variables
$\{\theta_j\}$ and $\{\phi_j\}$.

The random-phase Villain model can be treated in the same manner.
The results corresponding to (\ref{XY-E}),  (\ref{XY-gauge-correlation}),
(\ref{XY-inverse-correlation}) and (\ref{XY-correlation}) are,
under the condition $K=K_p$ and with the notation $T=K^{-1}$,
 \begin{eqnarray}
   && \left[ \langle e^{i(\theta_0 -\theta_1 -A_{01})} \rangle \right]
    =e^{-T/2} 
          \label{Villain-E} \\
  && \left[ \langle e^{i (\theta_0-A_{01}-A_{12}-\cdots
   -A_{r,r-1}-\theta_r)} \rangle \right] = e^{-rT/2}
           \label{Villain-gauge-correlation} \\
  &&  \left[ \frac{1}{\langle e^{i \sum_j a_j \theta_j}\rangle }
       \right] =1
                    \label{Villain-inverse-correlation} \\
   && \left[ \langle e^{i(\theta_0 -\theta_r)} \rangle \right]
   =\left[ \left|\langle e^{i(\theta_0 -\theta_r)} \rangle \right|^2\right]
                    \label{Villain-correlation}.
 \end{eqnarray}
Correlations between bond variables satisfy, similarly to
(\ref{XY-cos-correlation}) and (\ref{XY-sin-correlation}),
  \begin{eqnarray}
    &&\left[ \langle \cos (\theta_0-\theta_1-A_{01} )
    \cos (\theta_r-\theta_{r+1}-A_{r,r+1} ) \rangle \right]
   =  e^{-T}
        \label{Villain-cos-correlation} \\
  && \left[ \langle \sin (\theta_0-\theta_1-A_{01} )
    \sin (\theta_r-\theta_{r+1}-A_{r,r+1} ) \rangle \right]
   =0.  \label{Villain-sin-correlation}
 \end{eqnarray}
It can also be proved that the functional identity $P_m(z)=P_q(z)$
holds if $K=K_p$.
These relations for the Villain model agree with the corresponding
results for the $XY$ model in the low-temperature limit.

It is clear that the Nishimori line $K=K_p$
plays key roles in the determination of the system properties
also in the present continuous-variable system.
This line is expected to mark the crossover between the
pure ferromagnetic-like region and a randomness-dominated
region within the ferromagnetic phase.

\section{Spin wave theory}
The random-phase $XY$ model has often been analyzed by the spin
wave theory, particularly in two dimensions.
In the spin wave theory one expands the cosine interaction 
(\ref{XY-Hamiltonian}) to
second order of the argument, assuming that the argument is small
at low temperatures
  \begin{equation}
    \cos (\theta_i -\theta_j-A_{ij})\approx 1-\frac{1}{2}
      (\theta_i -\theta_j-A_{ij})^2.
        \label{SW}
  \end{equation}
It will be assumed that the quenched gauge variable $A_{ij}$
is also Gaussian-distributed with variance $\sigma \equiv K_p^{-1}$.

An important difference between the spin wave theory (\ref{SW}) and the
Villain model (\ref{Villain-BF}) is in periodicity.
The latter model is periodic in the variables $\theta_i, \theta_j$ and
$A_{ij}$ with period $2\pi$ as in the original
$XY$ model (\ref{XY-Hamiltonian})
whereas the spin wave theory (\ref{SW}) neglects periodicity.
Periodicity of the Hamiltonian (\ref{XY-Hamiltonian})
or the Boltzmann factor (\ref{Villain-BF}) is
directly related with the existence of vortices, which
leads to the Kosterlitz-Thouless transition in two dimensions
\cite{Kosterlitz-Thouless}.
It is then expected that the spin wave theory fails to
reproduce important characteristics of the
$XY$/Villain model.
Surprisingly this is {\em not} the case at least for
the relations we have derived above using gauge transformation
as shown below.

The spin-wave Hamiltonian is Gaussian, and everything can be evaluated
explicitly in any dimension.
The results corresponding to (\ref{Villain-E}) to
(\ref{Villain-sin-correlation}) are given below.
Note that the following relations are valid for any $T$ and $\sigma$
in contrast to the results of the
gauge theory which require $T=\sigma ~(K=K_p)$.
The spatial dimension is arbitrary.
  \begin{eqnarray}
       && \left[ \langle e^{i(\theta_0 -\theta_1 -A_{01})} \rangle \right]
    = \exp \left\{ (T-\sigma )G_{01} -\frac{\sigma}{2}\right\}
          \label{SW-E} \\
    && \left[ \langle e^{i (\theta_0-A_{01}-A_{12}-\cdots
   -A_{r,r-1}-\theta_r)} \rangle \right] 
     =\exp \left\{ (T-\sigma )G_{0r} -\frac{r\sigma}{2}\right\}
          \label{SW-gauge-correlation} \\
    &&  \left[ \frac{1}{\langle e^{i \sum_j a_j \theta_j}\rangle }
       \right] =\exp \left\{ \frac{T}{2}\sum_{i,j}a_i a_j G_{ij}
       -\frac{\sigma}{2}\sum_{i,j}a_i a_j G_{ij}\right\}
                    \label{SW-inverse-correlation} \\
   && \left[ \langle e^{i(\theta_0 -\theta_r)} \rangle \right]
       =e^{(T+\sigma )G_{0r}}
                   \label{SW-correlation1}\\
    &&\left[ \left|\langle e^{i(\theta_0 -\theta_r)} \rangle \right|^2\right]
        =e^{2TG_{0r}}
                    \label{SW-correlation2}.
  \end{eqnarray}
Here $G_{0r}$ is the lattice Green function.
Its explicit form is, for example for the two-dimensional square lattice,
  \begin{equation}
  G_{0r}=\frac{1}{N}\sum_k \frac{\cos \kv\cdot\rv -1}{4-2\cos k_1-\cos k_2}.
  \end{equation}
Correlations of bond variables can also be evaluated by the
spin wave theory:
  \begin{eqnarray}
    &&\left[ \langle \cos (\theta_0-\theta_1-A_{01} )
    \cos (\theta_r-\theta_{r+1}-A_{r,r+1} ) \rangle \right]
   =  \frac{1}{2} C_{r-}+\frac{1}{2} C_{r+}
        \label{SW-cos-correlation} \\
  && \left[ \langle \sin (\theta_0-\theta_1-A_{01} )
    \sin (\theta_r-\theta_{r+1}-A_{r,r+1} ) \rangle \right]
   =\frac{1}{2} C_{r-}-\frac{1}{2} C_{r+},
        \label{SW-sin-correlation}
 \end{eqnarray}
where $C_{r\pm }$ is, in the case of the square lattice,
  \begin{equation}
    C_{r\pm } =\exp \left\{ \frac{\sigma -T}{N}
    \sum_k \frac{(1-\cos k_1)(1\pm \cos \kv\cdot \rv)}{2-\cos k_1-\cos k_2}
 -\sigma \right\}.
  \end{equation}
All these relations (\ref{SW-E}) to (\ref{SW-sin-correlation})
reduce to the corresponding Villain results
(\ref{Villain-E}) to (\ref{Villain-sin-correlation})
if $T=\sigma$.
This is a surprising conclusion because it means that,
if $T=\sigma$, periodicity of the Villain model (or
vortex degrees of freedom) has no effects at all
on the quantities calculated above.
The spin wave theory gives the {\em exact} solution and {\em rigorous}
relations for the Villain model.
\section{Conclusion}
We have shown a number of exact/rigorous results for the
Ising spin glass and random-phase $XY$ and Villain models.
These models can be treated by the gauge theory,
a method exploiting gauge invariance of the Hamiltonian
or the Boltzmann factor.
The exact solution for the energy and rigorous relations for the
specific heat and correlation functions have been derived
under the condition of $K=K_p$.
The results for the $XY$ model have been shown to reduce to
those for the Villain model in the low-temperature limit.

We have used the spin wave theory to calculate
various correlation functions for the random-phase $XY$/Villain
models.
It has been found that the spin wave theory, which neglects
periodicity of the system, gives the exact expressions for the
quantities that have been treated by the gauge theory
for the Villain model
as long as the condition $K=K_p$ is satisfied.
Thus the effects of periodicity, or the vortex degrees of freedom,
completely disappear at least from these quantities if $K=K_p$.

It is of course expected that periodicity plays essential
roles in the determination of the explicit exact form of the correlation
function $[\langle e^{i(\theta_0 -\theta_r)} \rangle ]$
of the Villain model since, otherwise, the spin
wave theory would be valid for an arbitrarily high temperature,
implying that the low-temperature Kosterlitz-Thouless
(in two dimensions) or the ferromagnetic (in higher dimensions)
phase would extend to the infinite-temperature limit.
Nevertheless, the present conclusion of the exact relations
for the energy and other quantities by the spin wave theory
is never trivial, in consideration of the fact that
the Nishimori line $K=K_p ~(T=\sigma)$
seems to occupy no special status in the phase diagram
according to the renormalization-group treatment of the two-dimensional
Villain model \cite{Scheidl,Tang,Carpentier-LeDoussal}.
The renormalization group,
rather, predicts that the line defined by $T=2 \sigma$
separates the usual ferromagnetic region and the
vortex-freezing region within the ferromagnetic phase.
One possibility is that current renormalization group
treatments miss some important aspects of the system,
thus identifying the line $T=2 \sigma$ as the crossover line
instead of the line $T=\sigma$.
Further investigations are necessary to clarify this point.


\end{document}